# Fully passive monolithic silicon quantum photonic circuit for entangled photon-pair generation

*David E. Medina*[1*], *Paul J. Robin*[1], *Romain Dalidet*[2], *Sébastien* Tanzilli[2], Anthony Martin[2], Ség*olène Olivier*[3], *Quentin Wilmart*[3], *Laurent Vivien*[1], *Laurent Labonté*[2], *Carlos Alonso-Ramos*[1], *Eric Cassan*[1]

[1]*Centre de Nanosciences et de Nanotechnologies, Université Paris-Saclay, CNRS, 91120 Palaiseau, France*

[2]*Université Côte d'Azur, CNRS, Institut de Physique de Nice (INPHYNI), Parc Valrose, 06108 Nice Cedex 2, France*

[3] *CEA-LETI-Minatec, 17 rue des martyrs, Grenoble Cedex 09, France*

* Corresponding authors: eric.cassan@universite-paris-saclay.fr, laurent.labonte@univ-cotedazur.fr

Keywords: On-chip quantum optics, photon-pair source, silicon photonics, integrated pump filter, wavelength demultiplexing, low crosstalk, Bragg grating, photon counting, photon entanglement

**ABSTRACT:** Integrated quantum photonic circuits are a key enabling technology for the scalability of quantum information systems. Among the available platforms, silicon photonics offers an unrivalled capability for the large-scale integration of photonic components within compact footprints. However, the strong index contrast that enables ultra-compact silicon devices also makes them highly sensitive to fabrication imperfections. As circuit complexity increases, active tuning is generally required to maintain spectral alignment among the different components, leading to significant power consumption that ultimately limit scalability. Here, we demonstrate a fully integrated silicon quantum photon-pair source operating without active tuning of any component. The circuit combines photon-pair generation in a micro-ring resonator, pump rejection using Bragg filters, and signal/idler demultiplexing through modal add-drop filters with building blocks engineered to minimize sensitivity to fabrication variations. The resulting circuit achieves excellent experimental quantum performance. Coincidence rates up to 4000 counts $s^{-1}$ with coincidence-to-accidental ratios as high as 100 are obtained across the generated spectrum, while separate two-photon interference measurements yield raw visibilities exceeding 93% for individually selected ITU wavelength-channel pairs. By eliminating the need for active spectral tuning while maintaining high quantum performance, this work addresses a major bottleneck in the scaling of complex silicon quantum photonic circuits.

## 1 …|… INTRODUCTION

There is a growing demand to develop technologies capable that apply quantum mechanics principles to real-world applications, including quantum communications, distributed quantum computing, quantum sensing and metrology [1]. Among these, quantum networks are expected to play a central role, relying critically on efficient, scalable and deployable sources of entangled photon pairs. Integrated photonics has consequently emerged as one of the most promising platforms for generating, manipulating and detecting quantum states of light on a single chip, offering scalability, compactness and mechanical robustness while benefiting from mature microfabrication technologies [2,3]. Remarkable progress has recently been achieved across several integrated photonic platforms, including silicon nitride, thin-film lithium niobate, AlGaAs and hybrid technologies, significantly expanding the capabilities of integrated quantum photonics [4-6]. Among these complementary platforms, silicon-on-insulator (SOI) remains particularly attractive for implementing large-scale quantum photonic integrated circuits owing to its CMOS compatibility, high integration density and mature fabrication ecosystem. These advantages have enabled increasingly sophisticated quantum photonic devices, ranging from integrated entangled-photon sources to programmable quantum processors and multidimensional quantum state manipulation [7-9]. Silicon further provides an efficient third-order optical nonlinearity, enabling the generation of correlated and entangled photon pairs through spontaneous four-wave mixing (FWM) in compact waveguides and micro-ring resonators [10,11].

As integrated quantum photonic circuits become increasingly sophisticated, however, the challenge is no longer the realization of individual building blocks, but their reliable co-integration into a single manufacturable chip. Entangled photon-pair sources require several wavelength-selective functions, including photon-pair generation, pump rejection and wavelength demultiplexing, which operations critically depend on accurate spectral alignment. Fabrication-induced dimensional variations inevitably shift the resonance wavelengths of these components, making their simultaneous operation increasingly difficult as circuit complexity grows. Consequently, active thermal tuning has become the standard approach to compensate fabrication imperfections. Although highly effective, this solution requires continuous electrical power, dedicated control electronics and calibration procedures while introducing thermal crosstalk, ultimately limiting the scalability, energy efficiency and deployability of large quantum photonic circuits [2,8].

Considerable efforts have therefore focused on improving the individual building blocks required for integrated quantum photonics [12-14]. High-brightness silicon micro-ring resonators have demonstrated efficient photon-pair generation [10-11,15-17], integrated Bragg filters have achieved pump rejection exceeding 80 dB [18-20], and several wavelength-demultiplexing architectures based on ring resonators, Bragg gratings, Mach-Zehnder interferometers and arrayed waveguide gratings have demonstrated excellent filtering performance [21-24]. More recently, fully integrated quantum photonic circuits have demonstrated increasingly complex functionalities, including quantum frequency processing and multidimensional state manipulation [25]. However, while each of these functions can be individually optimized, integrating them into a cohesive high-performance quantum photonic circuit remains challenging. Beyond the performance of each individual component, the scalability of integrated quantum photonics ultimately relies on the spectral compatibility of all wavelength-selective building blocks after fabrication.

Here, we demonstrate a fully passive monolithic silicon quantum photonic circuit integrating the three essential functions of an on-chip entangled photon-pair source, namely photon-pair generation, pump rejection and wavelength demultiplexing. Rather than relying on active spectral correction, our approach is based on the co-design of passive building blocks engineered to remain spectrally compatible despite fabrication-induced variations. Implemented in a standard 300-nm silicon-on-insulator platform using a single-etch fabrication process, the circuit operates without active tuning of any optical component while remaining fully compatible with CMOS manufacturing. The fabricated device achieves coincidence-to-accidental ratios up to 100 across the generated spectrum, while raw Franson interference visibilities exceeds 93% over several International Telecommunication Union (ITU) wavelength channels. These results demonstrate that system-level co-design of fabrication-tolerant passive building blocks provides a practical route towards scalable, energy-efficient and manufacturable silicon quantum photonic circuits.

The proposed functional circuit, depicted in **Fig. 1**, is implemented in practice in a 300 nm-thick silicon-on-insulator platform with a 3 µm buried oxide layer, operating with transverse-electric (TE) polarized light near a wavelength of 1550 nm. Light is coupled into and out of the chip using fiber-chip grating couplers with an insertion loss level of 3.5 dB at 1550 nm wavelength for TE light polarization. The first stage consists of a dispersion-engineered silicon micro-ring resonator, exploiting resonant enhancement of light-matter interactions to maximize photon-pair generation via the FWM process in silicon. The optimized resonator exhibits a quality factor of approximately 150,000, operates at critical coupling, and provides a free spectral range (FSR) of ~100 GHz, matched to the ITU grid. This strong field enhancement increases the nonlinear interaction efficiency, enabling entangled photon-pair generation.

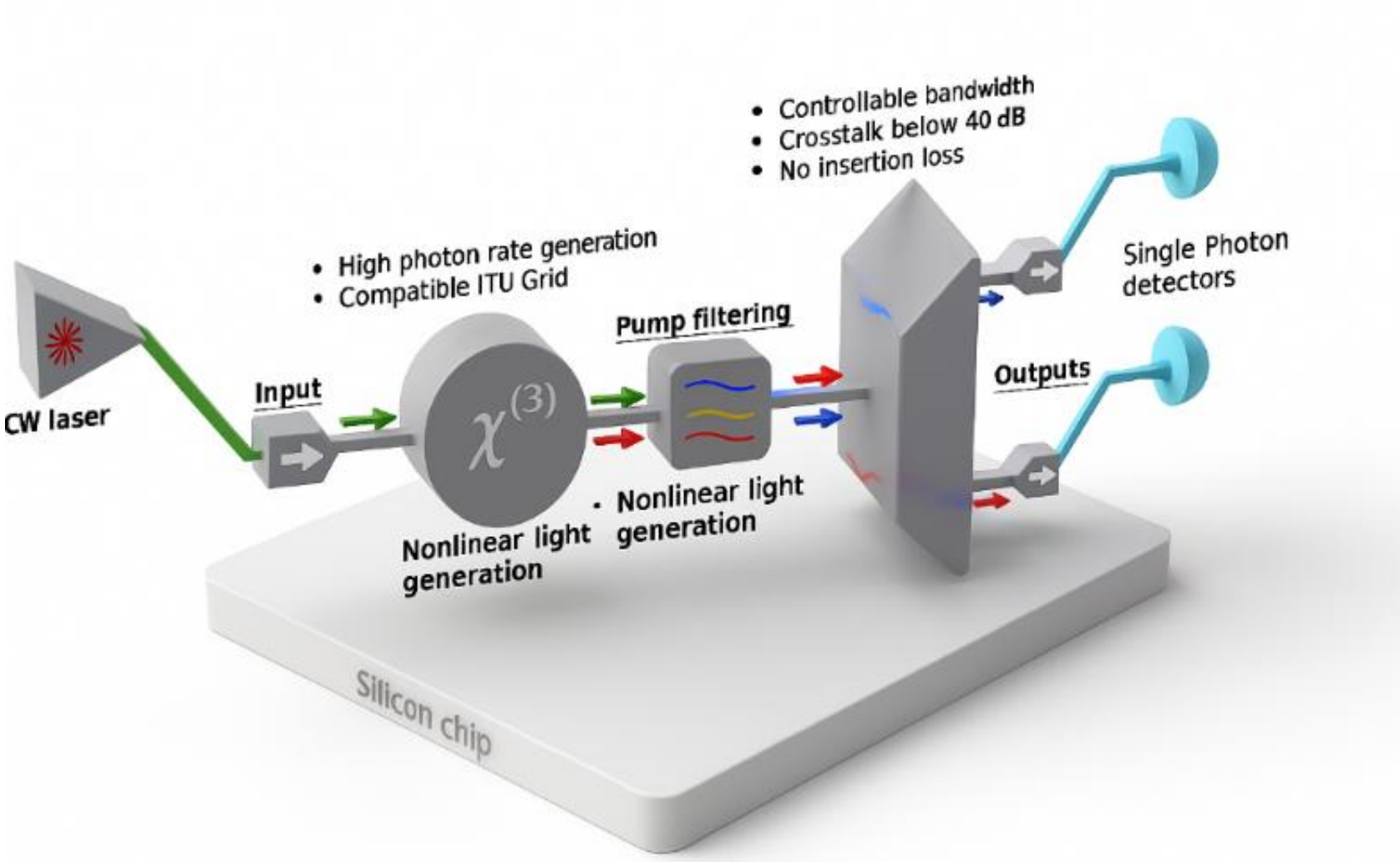

***Fig. 1.*** *Functional schematic of the realized entangled photon pair generator in a single fully-passive silicon photonic chip fed by a continuous wave tunable telecom wavelength (1.55µm; <2mW). Generated photon pairs are extracted in view of photon counting and quantum interference measurements, respectively.*

As discussed below, the pump power is optimized to balance efficient pair generation against degradation of the coincidence-to-accidental ratio (CAR) due to two-photon absorption (TPA) in silicon. The generated photons are subsequently routed to an integrated pump rejection stage based on the architecture introduced in [19,20]. This stage combines wavelength-selective suppression of the pump in the TE polarization with broadband attenuation of residual light in the transverse-magnetic (TM) polarization. The rejection filter has a 5 nm bandwidth, comprising multiple ring resonances. This design ensures that, despite fabrication-induced spectral misalignments between the ring and the rejection filter, at least one resonance always falls within the rejection band. The resulting device achieves a measured extinction ratio exceeding 60 dB. Finally, the signal/idler demultiplexing stage is implemented using a modal add-drop filter consisting of a multimode Bragg grating and a mode-selective directional coupler [26]. This architecture offers two key advantages: first, it enables flexible control of the signal and idler channel bandwidth through adjustment of the Bragg grating strength; second, it provides channel separation without requiring interferometric structures, thereby eliminating the need for active phase stabilization. The filter is designed with a 3 nm bandwidth for both signal and idler, enabling simultaneous collection of multiple channels and achieving an insertion loss below 1 dB and a measured extinction ratio of 40 dB. The complete circuit therefore integrates the three essential functions of an on-chip entangled photon-pair source, pair generation, pump rejection and wavelength demultiplexing, within a single monolithic silicon chip, without requiring active tuning of any optical component.

## 2 …|…KEY BUILDING BLOCKS

**Ring resonators**: The ring resonator geometry, schematically depicted in **Fig. 2a)** makes use of a pulley configuration for the coupling to the bus waveguide. In this configuration, the coupling between the ring and the bus waveguide is controlled by the gap and the angle $\boldsymbol{\vartheta}$. Ring resonators using this design achieve loaded quality factors ($Q$-factor) easily of more than 100k. The parameters for critical coupling were determined experimentaly by a thorough analysis of a sweep of parameters in a set of fabricated rings, leading to the results shown in **Fig. 2.b)**.

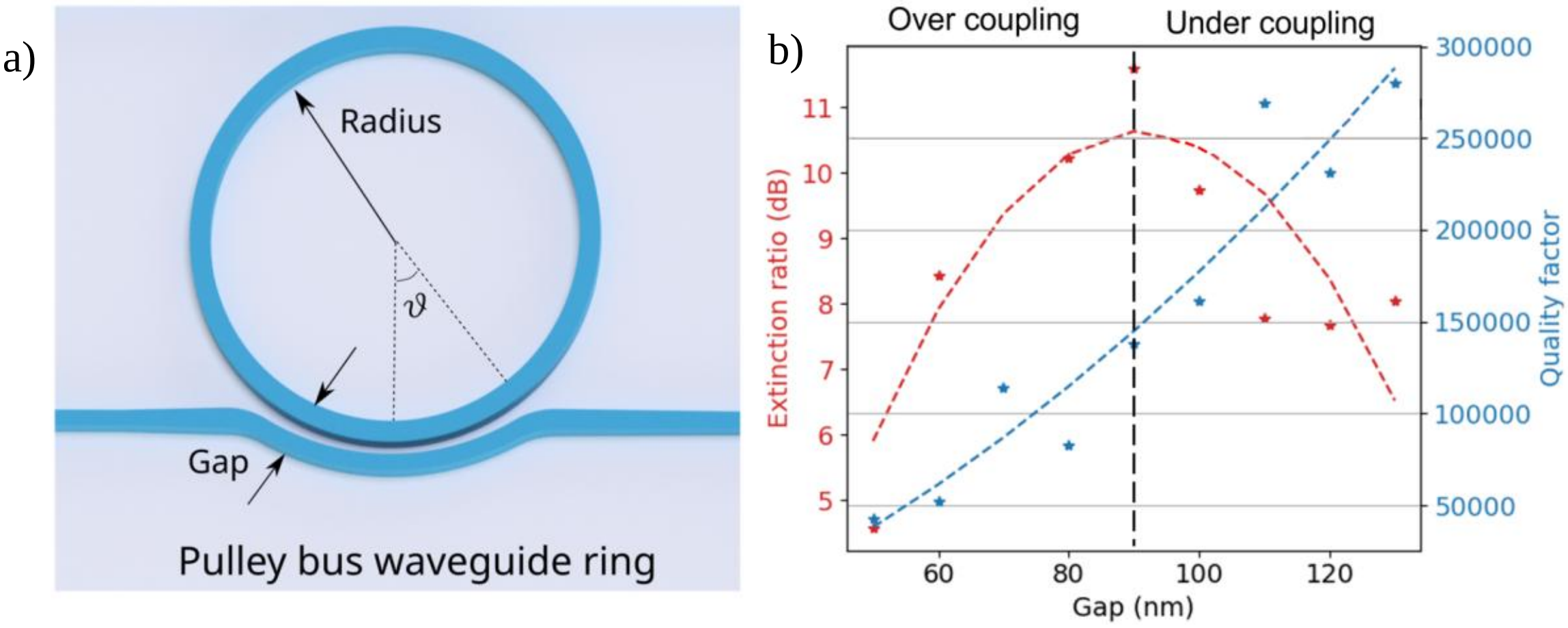

*Fig. 2.* ***a)*** *All pass ring resonator schematic highlighting the considered geometrical parameters: fabrication started from a SOI wafer with a silicon thin film thickness of 300nm on top of a 3 µm thick $SiO_2$ BOX and 600 nm wide SOI waveguides with air cladding were fabricated and characterized in TE-like light polarization.* ***b)*** *Quality factors and extinction ratio of the measured rings as a function of the gap for different fabricated components.*

Identifying the ring/bus waveguide coupling nature between under to over coupling further facilitated the quantification of the chips' quantum properties. A ring gap of ~90 nm was identified as matching the critical coupling condition. Slightly lower values (~80 nm), achievable with standard deep-UV lithography tools, could ensure a slightly over-coupled regime. This allowed fine-tuning of the pump efficiency for photon pair generation, while still favoring heralding efficiency through optimal photon extraction from the ring resonator**.** The optical pump signal was slightly tuned to probe a resonance near 1.55 µm wavelength, within the operational bandwidth of the grating couplers. This excited and enabled the collection of demultiplexed signal and idler resonances, typically corresponding to ITU channels with relative azimuthal numbers within ±10 of the pump resonance.

**Pump filter**: In spontaneous four-wave mixing (SFWM) experiments, the pump field is typically many orders of magnitude stronger than the generated signal and idler photons, posing significant challenges for their detection and characterization. A critical technical hurdle is thus the efficient suppression of this pump signal, as residual pump light can overwhelm the single-photon detectors and obscure the quantum correlations of interest. This issue is further compounded by the proximity of the signal and idler wavelengths to the pump, often separated by only a few nanometers, which complicates spectral filtering**.**

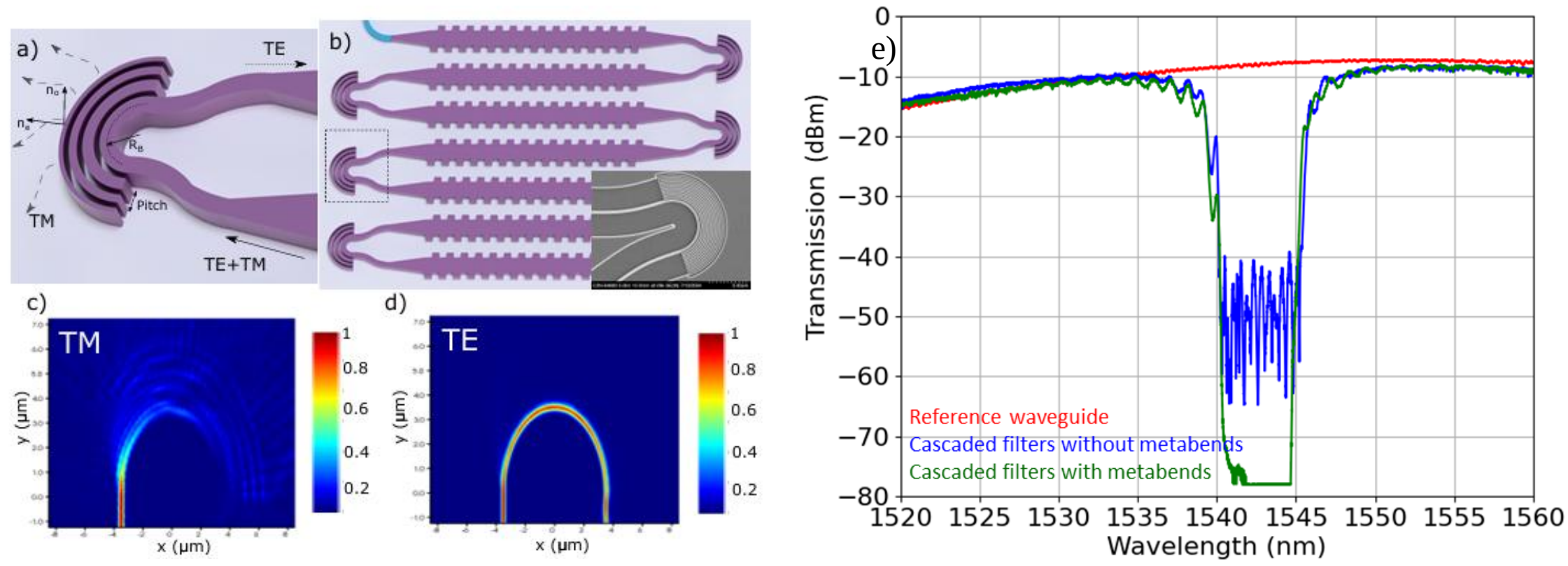

***Fig. 3.*** *Schematic of the implementation of subwavelength metabend structures into the whole cascaded Bragg filter.* ***a)*** *3D schematic representation of the metabend ridges transmitting TE waves while radiating TM polarized light;* ***b)*** *Overview of the multisection cascaded Bragg filter;* ***c)*** *and* ***d)*** *FDTD computed light propagation at the pump wavelength (1553nm);* ***e)*** *Comparison of the transmission spectra of a simple cascaded multimode Bragg filter with 10 sections with cascaded filters, without and with ridge waveguide metabends.*

High-extinction-ratio pump rejection on the order of 100 dB is typically required to prevent detector saturation and ensure reliable single-photon counting [4,18]. Additionally, the experimental setup must address photonic noise arising from scattered or leaked pump light, which can degrade the signal-to-noise ratio and spoil quantum interference measurements. Achieving such stringent pump suppression demands advanced filtering techniques, such as cascaded Mach-Zehnder interferometers, cascaded Bragg gratings, integrated resonators filters [4,24]. Our implementation of the pump filter is based on broken coherence cascaded waveguide Bragg filters that yielded 80 dB measured rejection with 220-nm-thick silicon waveguides [19]. The achievable rejection in the 300-nm-thick Si technology is limited to ∼ 40-50 dB by residual light coupling in the transverse magnetic (TM) light polarization. However, the implementation of TM-filtering metabends [27] further enabled experimental rejection exceeding 60 dB by inducing additional losses for TM-polarized waves [20, 28]. The implementation of both metabend filters and cascaded Bragg section in our platform of interest is shown in **Fig. 3.a)-d)**. Through several fabrication and test cycles, the filter stage was optimized to improve rejection and reduce insertion losses. We achieved >70 dB rejection and <3 dB optical insertion loss at the wavelengths of interest, limited only by the noise floor of the experimental setup (see **Fig. 3e**).

**Wavelength demultiplexers of generated photon pairs**: Separating the signal and idler photons into different waveguide channels while achieving the highest possible extinction ratio and maintaining minimal insertion losses (typically below 1 dB) was next considered. Considering the approaches developed so far in this purpose [22], including ring resonators [14,15], Mach–Zehnder interferometers [22,24], Bragg gratings [23], and arrayed waveguide gratings (AWGs) [21], a modal add-drop (MAD) demultiplexer scheme was considered in this purpose. This device consists of a multimode Bragg grating combined with a mode-selective directional coupler [26], as schematically illustrated in **Fig. 4a**. The input waveguide is first connected to a mode-selective directional coupler, with its through port subsequently linked to an asymmetric Bragg grating. The directional coupler is designed to achieve phase matching between the first-order mode of the lower waveguide and the fundamental mode of the upper waveguide. The asymmetric Bragg grating is engineered to reflect light injected in the fundamental mode while converting it into the first-order mode within the rejection bandwidth, similarly to the pump rejection filter discussed in the previous section.

When light in the fundamental mode is injected into the input waveguide, it passes through the asymmetric coupler without coupling to the upper waveguide, since no phase matching exists for the fundamental mode of the lower waveguide. Light then reaches the Bragg grating, where, within the rejection bandwidth, it is converted into the first-order mode and reflected back toward the asymmetric coupler. In this case, phase matching occurs with the fundamental mode of the upper waveguide, enabling efficient coupling into the drop waveguide.

This architecture enables flexible control of the signal and idler channel bandwidths through adjustment of the Bragg grating strength. In addition, it provides channel separation without relying on interferometric structures, which are typically required in Bragg-based demultiplexers [23]. Since no interferometric effect is involved, the architecture also eliminates the need for active phase stabilization. Two such filters can be cascaded to separately extract the signal and idler photons. In each case, the asymmetric Bragg grating filter is designed for the specific wavelength corresponding to either the signal or idler photons. Beyond selecting the desired signal and idler photons, these devices also provide additional pump suppression, since the pump wavelength lies within the rejection bands of both the signal and idler filters.

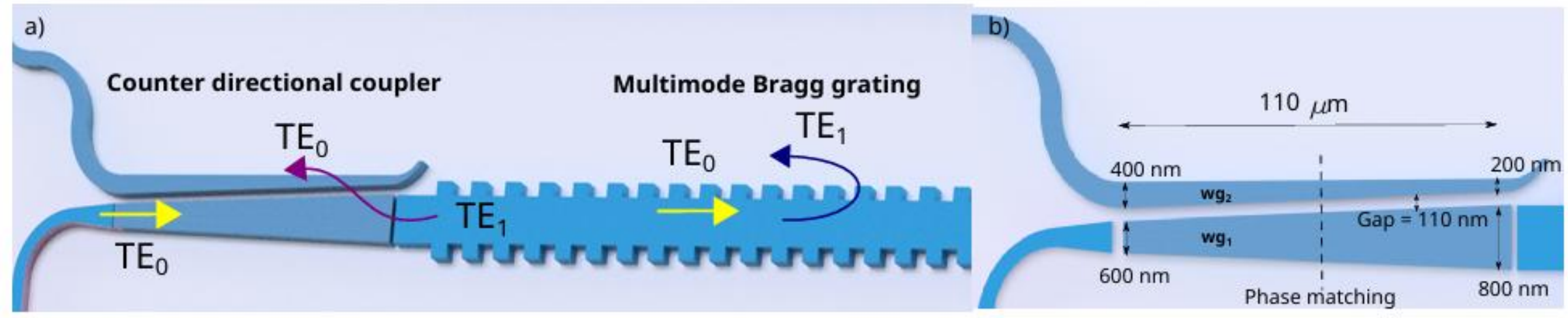

***Fig. 4**. **a)** Schematic of the proposed modal add coupler device, in which the TE1 reflected back light from the multimode Bragg grating couples into $TE_0$ in a counter directional coupler for a central wavelength of interest, **b)** Geometrical parameters optimized for the counter directional coupler that comply the mode phase matching conditions for backward light $TE_0$ to $TE_1$ mode transfer. These specific values linked to the 300 nm silicon thin SOI wafer and to the 600 nm wide SOI waveguides considered could be adapted to other photonic platforms.*

The simulated and experimental responses of the asymmetric Bragg grating are shown in **Fig. 5(a)** and **(b)**, respectively. The drop-port response exhibits pronounced ripples on both sides of the passband, significantly limiting the off-band rejection. This behavior is a well-known limitation of uniform Bragg grating filters. To mitigate the resulting crosstalk, an apodization scheme was implemented to suppress the sidelobes outside the passband. To this end, a transition section was introduced between the input/output waveguides and the Bragg grating, providing a gradual variation of the grating strength. Conventional Bragg apodization techniques typically rely on modifying the duty cycle and/or corrugation width to control the coupling strength. However, the implementation of weak gratings using these approaches often requires feature sizes that are incompatible with standard fabrication processes. Furthermore, variations in these parameters modify the effective index of the Bloch mode, resulting in a shift of the Bragg wavelength. Compensating for this effect requires chirping the grating period. Since fabrication imperfections can induce small dimensional deviations that alter the Bloch mode index, the required period compensation becomes highly sensitive to fabrication errors.

We implemented an alternative apodization strategy that overcomes these limitations. In this scheme, the grating strength is controlled through the relative lateral shift between the teeth on opposite sides of the grating. A shift of half a period maximizes coupling to the first-order mode, and thus the grating strength, whereas a zero shift, corresponding to a symmetric structure, minimizes it. This approach enables the realization of very weak strength gratings without reducing the minimum feature size. Moreover, varying the relative shift has a negligible impact on the Bloch mode index, allowing the transition to be implemented with a constant grating period.

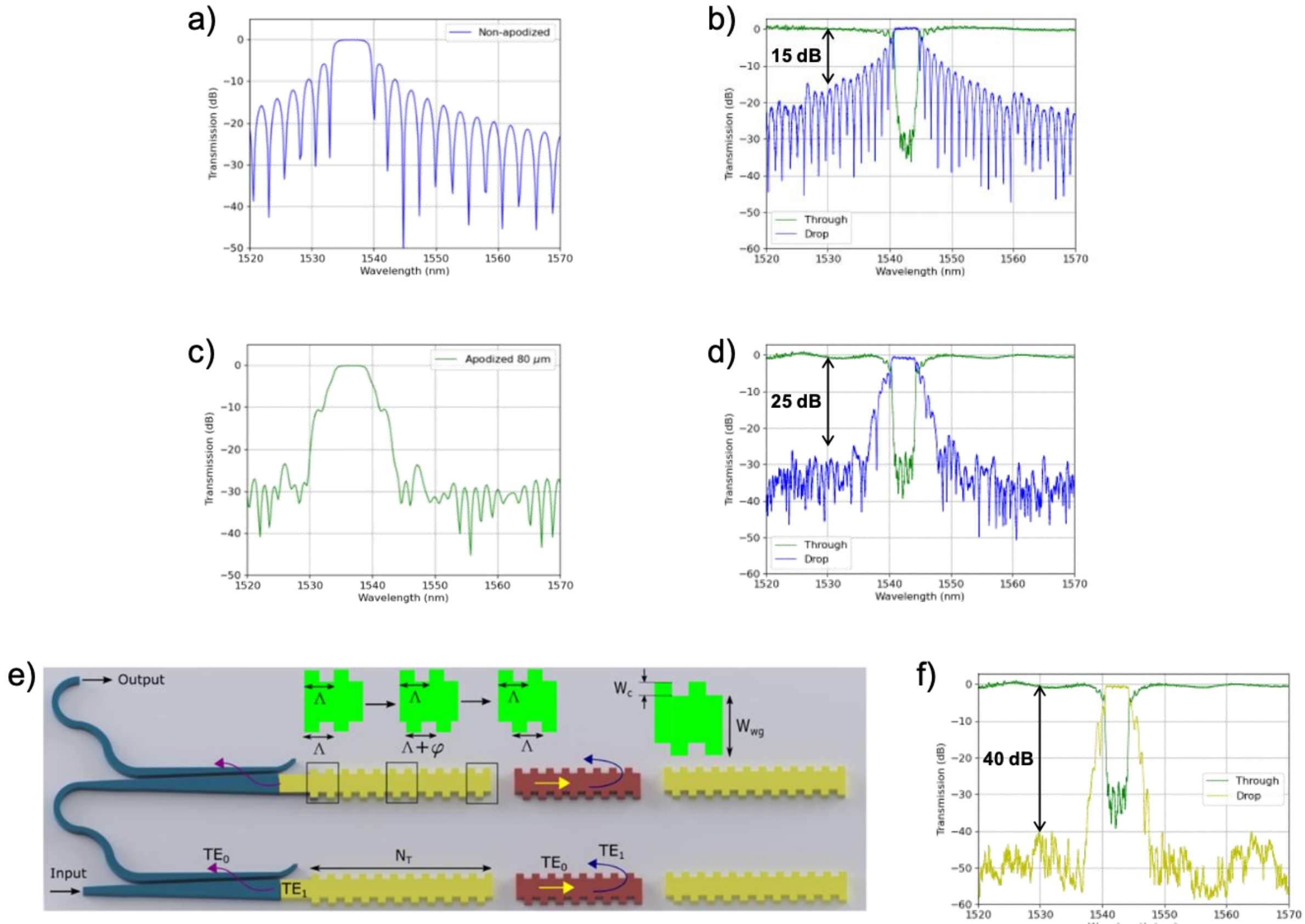

***Fig. 5**. Modal add-drop (MAD) filters. For a uniform grating: **(a)** Simulated drop-port transmission spectrum obtained by FDTD simulations. **(b)** Measured through- and drop-port transmission spectra. The slight wavelength shift with respect to the simulated response can be attributed to geometrical deviations introduced by lithography and etching. For an apodized grating with an 80 µm-long apodization section: **(c)** Simulated drop-port transmission spectrum obtained by FDTD simulations. **(d)** Measured through- and drop-port transmission spectra. **(e)** Layout of the cascaded MAD filter composed of two apodized sections. The inset illustrates the grating geometry used for apodization. **(f)** Measured transmission spectrum of the cascaded apodized MAD filter.*

The proposed transition was realized by linearly varying the relative tooth shift from zero to half the grating period ($\Lambda/2$) along the Bragg structure, as illustrated in **Fig. 5(e)**. The simulated and measured responses of the resulting apodized filter are shown in **Fig. 5(c)** and **(d)**, respectively. The device exhibits a passband centered at 1542 nm, negligible insertion loss, and an off-band rejection of 25 dB.

To further enhance the rejection, two MAD filters were cascaded, as shown in **Fig. 5(e)**. This configuration achieved an off-band suppression exceeding 40 dB while maintaining an insertion loss below 1 dB (see **Fig. 5f**). Such high rejection effectively minimized crosstalk between the signal and idler channels while providing additional pump suppression.

## 3. …|… PHOTON-PAIR GENERATOR

The building blocks previously described were combined to realize the complex circuit depicted in **Fig. 6**, implementing an entangled photon-pair generator with integrated pump-rejection and signal-idler demultiplexing. In this circuit, a ring resonator generates a comb of entangled-photon pairs. The remaining strong pump signal is rejected in a cascaded multi-mode coherency-broken Bragg grating, and signal and idlers are multiplexed in by the use of two MADs composed by two cascaded sections each. It is important to note that our approach uses only passive structures - no thermal heaters or active elements were used to align the optical resonance of the ring, pump-rejection or signal/idler demultiplexer.

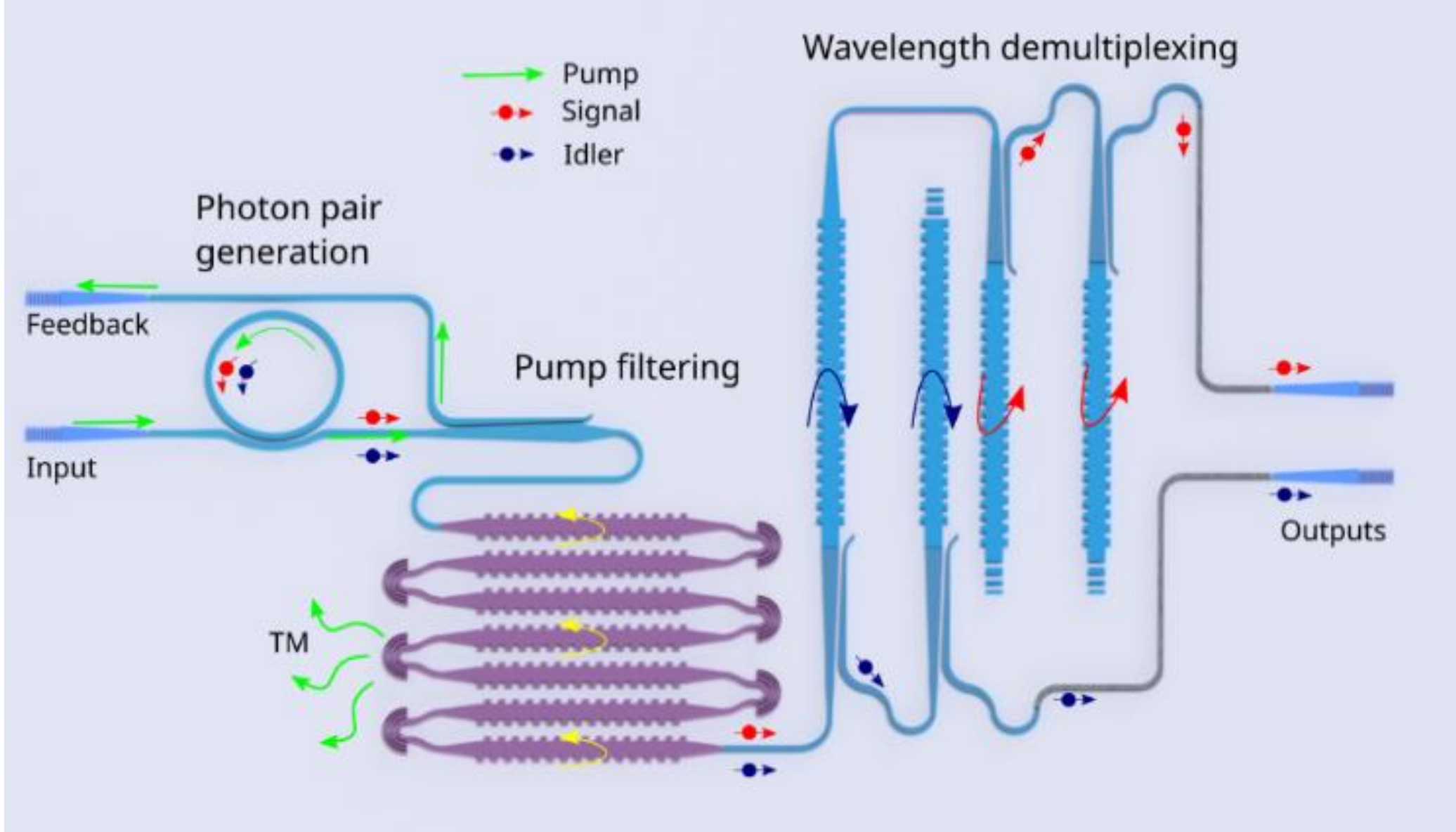

***Fig. 6**. Schematic view of the proposed photon-pair generator circuit. Pump, signal, and idler photons are illustrated by green, red, and blue arrows respectively. The two outputs highlighted on the right side represent the output ports for characterizing the photon pairs generated and their quantum correlations, while a single input signal feeds the circuits (the pump signal, the feedback port serving only to cross-check the bandwidth center of the pump filter stage).*

The measured transmission spectra of both demultiplexed idler and signal channels are shown in **Fig. 7a**. These two demultiplexed channels were designed to be shifted by a few nanometers from the pump filter's resonance, with a 5 nm bandwidth to capture 4-6 FSR resonances of the ring resonator, enhancing photon generation. For comparison, the transmission spectrum of a reference waveguide is also plotted, illustrating a total input-output loss of approximately 7.5 dB for the grating couplers. The additional insertion loss level of the signal and idler channels was mainly associated to the pump-rejection filter (2-3 dB) and the MAD demultiplexer (0.5-1 dB). The remaining losses were attributed to light waveguide propagation. Additionally, the setup used a single input fiber and a fiber array for output, with the fiber array introducing an additional 1-2 dB insertion loss compared to single fibers due to positioning limitations.
The measured filter rejection level was limited to 60 dB due to the noise floor of the measurement system. Optical characterization with an optical spectrum analyzer (OSA) refined this value to 80 dB for the combined filter and MAD actions.

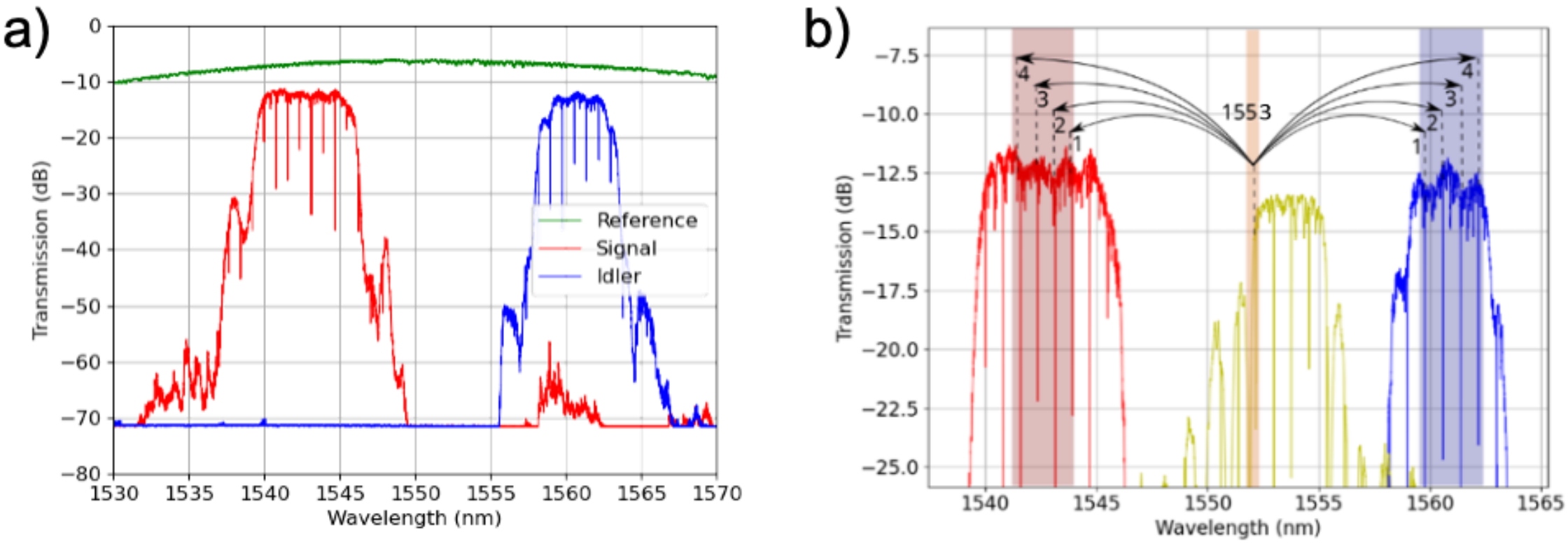

*Fig. 7. **a)** Transmission spectra of the demultiplexed idler/signal channels after passing through the three stages (entangled photon pair generation, pump filtering, and WDM), demonstrating the device's capability to efficiently separate signal and idler photons with high fidelity. **b)** Transmission spectra of the feedback port (yellow), signal (red) and idler (blue), illustrated the selected resonance for pump, with wavelength near 1553 nm, and the channels selected for signal and idler, conform to ITU grid specifications.*

After the classical circuit characterization, we selected an appropriate pump resonance to maximize simultaneous photon-pair generation in both channels. As shown in **Fig. 7b**, with the pump tuned to 1553 nm (aligned with ITU grid channel 30), four ring resonances were available for photon-pair generation, with each channel meeting energy conservation requirements, covering ITU grid channels 20-23 for the idler and 37-40 for the signal photons. The total insertion loss for the demultiplexed signal and idler photons was measured at 8 dB for the signal and 9 dB for the idler channels.

**Time correlation measurements:** To assess the performance of the integrated photon-pair source, photon time-correlation measurements were first performed using the experimental setup shown in **Fig. 8**. A continuous-wave tunable laser was spectrally cleaned using a Yenista XTM-50 filter providing approximately 50 dB rejection of amplified spontaneous emission before being coupled into the chip through a TE grating coupler. The chip temperature was actively stabilized at 25 °C using a Peltier stage, while a polarization controller ensured excitation of the fundamental TE optical mode. At the chip output, the signal and idler photons were collected using a fiber array and independently filtered by two tunable bandpass filters.

These filters simultaneously selected the wavelength channels of interest and removed the residual pump light. Combined with the on-chip pump rejection exceeding 80 dB, their additional rejection of approximately 30 dB was sufficient to suppress the remaining pump power before detection with superconducting nanowire single-photon detectors (SNSPDs).

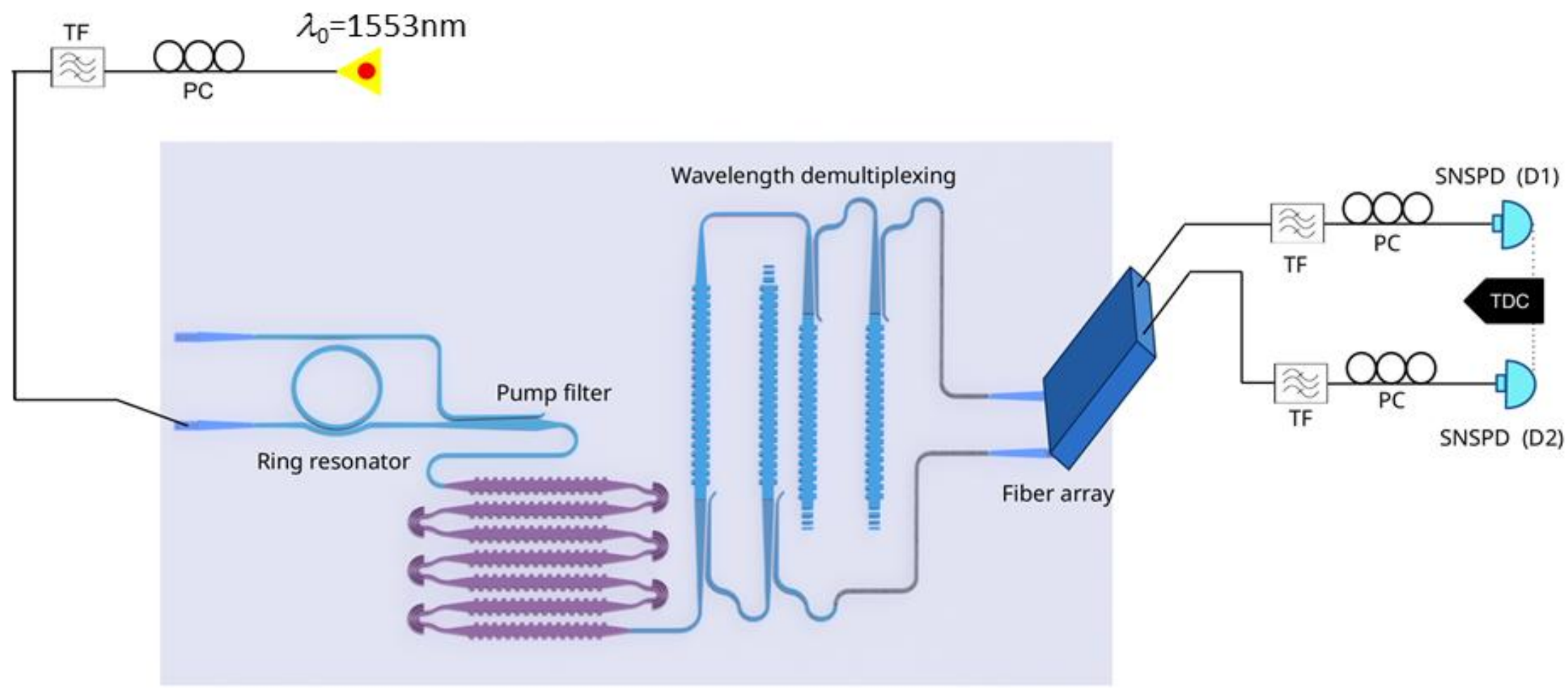

***Fig. 8***. *Schematic view of setup used for time correlation experiments. The entire setup uses single mode fibers (SMF) for the experiments.*

**Fig. 9(a)** presents the measured coincidence histogram obtained by collecting all resonances within the signal and idler spectral bands. The corresponding photon-pair generation rate and coincidence-to-accidental ratio (CAR) as a function of the on-chip pump power are shown in **Fig. 9(b)**. Both datasets are simultaneously described using a common phenomenological model in which the detected pair rate is expressed as $R(P_p) = aP_p + bP_p{}^2$, while the CAR follows $\mathrm{CAR}(P) = c/[R(P) + d]$, with $P_p$ the pump power. The fit yielded $a = 615\ s^{-1}mW^{-1}$, $b = 2223s^{-1}mW^{-2}$, $c = 4.55 \times 10^5 s^{-1}$, and $d \simeq 2.7 \times 10^{-16} s^{-1}$. The dominant quadratic coefficient confirms that the detected photon-pair rate is primarily governed by the SFWM process, while the comparatively small linear contribution most likely reflects spurious non-linear affect as Raman. Moreover, the negligible value of $d$ indicates that the CAR is almost entirely determined by the detected pair rate, with no measurable contribution from a pump-independent background, thus evidencing the very low level of parasitic noise in the source. This model therefore directly links the CAR degradation to the increase in the photon-pair generation rate through a common set of physically meaningful parameters. As expected, increasing the pump power enhances the SFWM efficiency, leading to a higher photon-pair generation rate, while the CAR gradually decreases due to the increasing probability of multiple-pair emission. An on-chip pump power of 1.2 mW provides an excellent compromise between source brightness and quantum performance, yielding a photon-pair generation rate of approximately $4 \times 10^3 \text{counts}^{-1}$ while maintaining a CAR above 100.

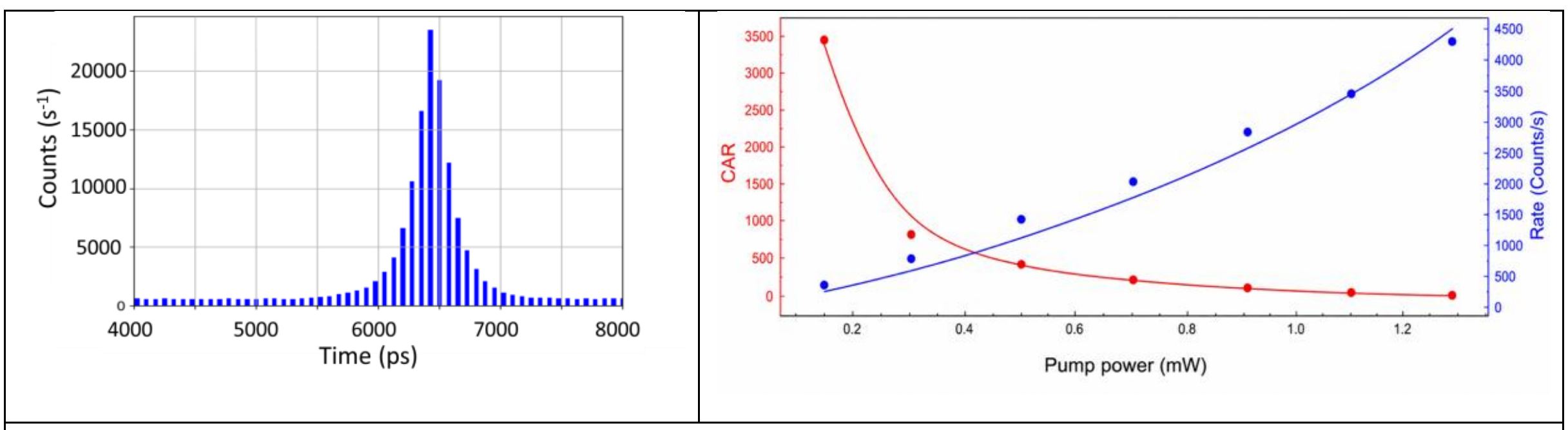

***Fig. 9*****.** **a)** *Photon arrival histogram for pump power in the input waveguide of 1.2 mW*. **b)** *Coincidence-to-accidental ratio (CAR) (red, left axis) and photon-pair generation rate (blue, right axis) as a function of the injected on-chip pump power. Solid lines are obtained from a simultaneous global fit.*

### Energy-time entanglement analysis:

The quality of the generated entangled photon pairs was subsequently assessed through energy-time interference measurements using a folded free-space Franson interferometer. In contrast to the previous measurements, where all generated wavelength channels were collected simultaneously, the interference characterization was performed independently for each signal-idler ITU channel pair. Raw (net) two-photon interference visibilities of 95.55 ± 2.98% (96.56 ± 2.99%), 97.13 ± 2.36% (97.13 ± 2.36%), 97.01 ± 2.31% (97.57 ± 2.33%), and 93.19 ± 2.86% (93.65 ± 2.87%) were obtained for the four investigated ITU wavelength-channel pairs, as shown in **Fig. 10**. The consistently high visibilities demonstrate the excellent quality of the generated energy-time entangled states and confirm that the complete integrated circuit, including the photon-pair source, the pump rejection stage, and the signal/idler wavelength demultiplexers, preserved the quantum coherence of the generated photons. Moreover, the close agreement between the raw and net visibilities indicates that only a negligible level of background noise was introduced by the integrated circuit, highlighting the effectiveness of the on-chip pump rejection and the low-loss wavelength demultiplexing architecture. Together, these results demonstrate that high-performance quantum operation can be achieved without requiring any active tuning of the photonic circuit. The residual reduction in visibility is likely dominated by the accumulation of detector dark counts during the 50 s integration time, rather than by imperfections of the integrated photonic circuit itself.

Beyond validating the spectral uniformity of the source, the simultaneous operation of multiple ITU wavelength-channel pairs highlights the potential of the proposed architecture for frequency-multiplexed quantum networking. A single integrated chip installed at a network node could generate broadband entangled photon pairs and distribute them over conventional dense wavelength division multiplexing (DWDM) telecommunication infrastructure, while wavelength demultiplexers located at the user side would select the appropriate ITU channel pair for each quantum communication link [29]. Such an architecture would enable multiple entanglement-based BBM92 QKD connections to operate simultaneously from a single integrated source [30]. Importantly, the circuit could be used directly without active spectral tuning while satisfying the stringent visibility requirements of entanglement-based QKD protocols. Indeed, raw two-photon interference visibilities exceeding 90% correspond to quantum bit error rates below 5%, ensuring secure operation without the need for background-noise subtraction.

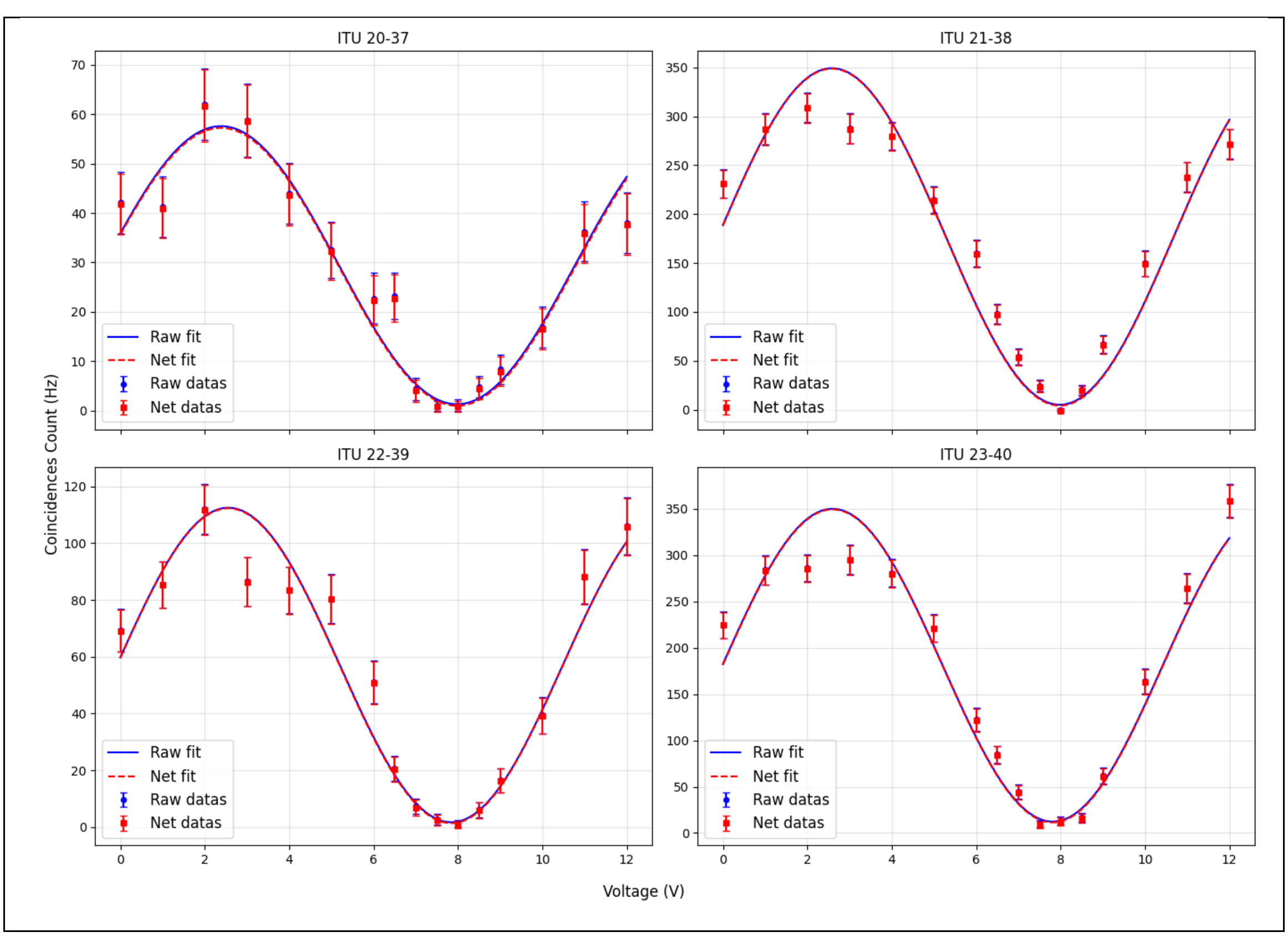

***Fig. 10.*** *Two-photon Franson interference fringes measured for the four demultiplexed ITU channel pairs (20–37, 21–38, 22–39, and 23–40). Coincidence counts were recorded as a function of the phase applied to the folded Franson interferometer by varying the voltage applied to the piezoelectric actuator inside the interferometer, with an integration time of 20 mHz per point. Blue (red) squares correspond to the raw (net) coincidence counts, with error bars given by the corresponding Poissonian statistics. The solid blue curves represent fits to the raw data, while the dashed red curves are sinusoidal fits to the net data used to extract the interference visibility.*

**Table 1** compares representative demonstrations of integrated silicon photon-pair sources reported over the past decade. Beyond conventional quantum metrics, such as photon-pair generation rate, CAR, and raw two-photon interference visibility, it also compares key system-level characteristics including the level of photonic integration, passive operation, on-chip pump rejection, and WDM compatibility. Rather than ranking the different approaches according to a single figure of merit, this comparison highlights the technological trade-offs between circuit complexity, scalability, and quantum performance. Earlier demonstrations generally focused on optimizing individual building blocks or maximizing a particular performance metric, such as CAR or interference visibility, often relying on partially integrated architectures or active tuning of wavelength-selective components [15,17,18,31]. More recent works have demonstrated increasingly sophisticated photonic circuits, but at the expense of additional control complexity associated with resonant devices requiring thermal stabilization [4,9,32,33]. In contrast, the present work simultaneously integrates the three essential functions of an entangled photon-pair source within a single monolithic silicon chip operating without active spectral tuning. Although our objective is not to maximize a single performance metric, the circuit nevertheless achieves competitive quantum performance. This combination of passive operation, high integration level, WDM compatibility, and competitive quantum performance represents, to the best of our knowledge, the first demonstration of a fully passive monolithic silicon quantum photonic circuit integrating all the essential functions required for entangled photon-pair generation.

| | System-level integration | | | | Quantum performance | | |
|---|---|---|---|---|---|---|---|
| **Reference** | **Pump rejection** | **Single chip** | **Passive** | **WDM compatible** | **Pair rate (Hz)** | **CAR** | **Raw visibility (%)** |
| [15] | Bragg filters | No | Yes | Yes | <1 | 45 | NA |
| [18] | Cascaded MZIs | No | No | No | <1 | 2.5 | NA |
| [31] | Cascaded MRRs | Yes | No | No | 100–500 | 15 | 81 |
| [17] | Coupled MRRs | Yes | No | No | 50–100 | 100–200 | NA |
| [11] | Cascaded Bragg filters | Yes | Yes | No | 100–200 | 67 | 98 |
| [32] | Cascaded CDCs | Yes | Yes | No | <1 | 27 | NA |
| [9] | Cascaded MZIs | Yes | No | Yes | 10000 | NA | 98.5 |
| [33] | Cascaded MRRs | Yes | No | Yes | 40 | 5 | NA |
| [4] | MZIs and MRRs | Yes | No | Yes | NA | 1000 | 99.5 |
| **This work** | **Cascaded Bragg filters** | **Yes** | **Yes** | **Yes** | **4000** | **100** | **93–97** |

***Table 1***. *Comparison of representative integrated silicon photon-pair sources reported at telecom wavelengths, including both quantum performance metrics and key photonic integration features.*

## 4. …|…DISCUSSION AND CONCLUSION

The realization of large-scale quantum photonic systems requires integrated platforms that simultaneously provide high quantum performance, low power consumption, and compatibility with scalable manufacturing. While silicon photonics offers unmatched integration density and mature CMOS fabrication processes, scaling quantum photonic circuits increasingly relies on the ability to co-integrate multiple wavelength-selective functions without introducing excessive system complexity.

In this work, we have demonstrated a fully integrated silicon quantum photonic circuit operating without active tuning of any component. By combining fabrication-tolerant photon-pair generation, pump rejection, and signal/idler demultiplexing within a single monolithic chip, we achieve photon-pair generation rates of up to 4000 counts $s^{-1}$ with coincidence-to-accidental ratios as high as 100. In separate two-photon interference measurements, raw visibilities exceeding 93% are obtained for individually selected ITU wavelength-channel pairs. The high interference visibilities reported here demonstrate that, once a given ITU channel pair is selected using standard DWDM filtering, the complete on-chip photonic circuit preserves the quantum coherence of the generated photon pairs without requiring any active tuning. Such frequency-multiplexed integrated sources constitute promising building blocks for future centralized entanglement distribution nodes serving multiple QKD users over existing telecom DWDM networks, where standard telecom filters can naturally perform the wavelength selection at the receiver side.

More generally, this work highlights the importance of system-level co-design in integrated quantum photonics. Rather than optimizing each building block independently, the proposed approach ensures that the spectral responses of all wavelength-selective components remain compatible after fabrication, thereby eliminating the need for active spectral correction. As integrated quantum photonic circuits continue to increase in complexity, reducing the dependence on active tuning is expected to become increasingly important, not only to decrease power consumption but also to simplify circuit control and improve scalability.

Looking ahead, the remaining building blocks required for fully integrated quantum photonic systems have already reached a high level of maturity. Hybrid integration techniques now enable the incorporation of III-V semiconductor lasers onto silicon-on-insulator (SOI) platforms through established wafer- and die-bonding processes [34]. In parallel, recent demonstrations of room-temperature germanium-silicon (GeSi) single-photon avalanche diodes (SPADs) provide a promising route towards CMOS-compatible on-chip single-photon detection without the need for cryogenic cooling [35]. Together, these developments indicate that all the essential components, photon sources, quantum photonic circuits, and single-photon detectors, could be integrated within a common silicon platform.

The results presented here therefore address one of the main bottlenecks limiting the scalability of silicon quantum photonics: the dependence on active spectral tuning. More broadly, they demonstrate that fabrication-tolerant system-level co-design enables high-performance quantum photonic functions to be co-integrated within a fully passive silicon platform. We believe that this approach provides a practical route towards large-scale, energy-efficient, and manufacturable silicon quantum photonic systems fully compatible with existing CMOS fabrication processes and telecommunications infrastructure.

**Acknowledgements**

SPHIFA ANR project, all teams involved at INPHYNI, C2N, and CEA-LETI research units. This work was also supported by the RENATECH network.

**Conflict of Interest**

The authors declare no conflict of interest.

**Data Availability Statement**

The data that support the findings of this study are available from the corresponding author upon reasonable request.